\documentclass[%
 reprint,
superscriptaddress,
 amsmath,amssymb,
 aps,
 prl,
]{revtex4-2}

\usepackage{graphicx}
\usepackage{dcolumn}
\usepackage{bm}
\usepackage{xcolor}
\DeclareMathOperator{\argminG}{arg\,min} 

\begin{document}

\title{Single-shot polarimetry of vector beams by supervised learning} 

\author{Davide Pierangeli}
\email{davide.pierangeli@roma1.infn.it}
\affiliation{Institute for Complex System, National Research Council (ISC-CNR), 00185 Rome, Italy}
\affiliation{Physics Department, Sapienza University of Rome, 00185 Rome, Italy}
\affiliation{Research Center Enrico Fermi (CREF), 00184 Rome, Italy}

\author{Claudio Conti} 
\affiliation{Physics Department, Sapienza University of Rome, 00185 Rome, Italy}
\affiliation{Institute for Complex System, National Research Council (ISC-CNR), 00185 Rome, Italy}
\affiliation{Research Center Enrico Fermi (CREF), 00184 Rome, Italy}

\begin{abstract}
States of light encoding multiple polarizations - vector beams - offer unique capabilities in metrology and communication.
However, their practical application is limited by the lack of methods for measuring many polarizations in a scalable and compact way.
Here we demonstrate polarimetry of vector beams in a single shot without any polarization optics.
We map the beam polarization content into a spatial intensity distribution through multiple light scattering and  
exploit supervised learning for single-shot measurements of multiple polarizations.
The method also allows us to classify beams with an unknown number of polarization modes, 
a functionality missing in conventional techniques.
Our findings enable a fast and compact polarimeter for polarization-structured light,
a universal tool that may radically impact optical devices for sensing, imaging, and computing.
\end{abstract} 

\maketitle

Generating, manipulating, and detecting optical states of polarization (SOP) is 
of paramount importance in many areas such as optical communication~\cite{Damask2005}, sensing~\cite{Tyo2006},
microscopy~\cite{Pierangelo2013}, and quantum information and computation~\cite{Crespi2011}.
While progress in material growth and nanotechnology are enabling advances in active polarization control~\cite{Zayats2017, Capasso2017, Atwater2021},
the measurement of light polarization remains limited by its intrinsic vectorial nature.
Complete determination of a single SOP needs at least four individual measurements,
each projecting the state on a distinct vector~\cite{Azzam2016, Tyo2002, Kurtsiefer2006}.
Conventional polarimetry methods replicate in time or space the polarization analyzer, 
which results in bulky optical setups, or in costly, compact polarimeters based on metasurfaces ~\cite{Brongersma2012, Pors2015, Bai2019, Mueller2016, Rubin2018, Zhang2019, Faraon2018, Capasso2019}.
While the need for several measurements is still affordable for light of uniform polarization,
it becomes a serious issue for beams with a spatial polarization structure.

Light with non-uniform polarization across the transverse plane exhibits non-separable correlations between polarization and spatial modes \cite{Aiello2015}. These vector beams recently have disclosed unique potentials in metrology \cite{Dambrosio2013, Marquardt2015, Fang2021},
communication \cite{Zhu2021, Forbes2022}, optoelectronics \cite{Sederberg2020}, optomechanics \cite{Brasselet2021}, 
and quantum information \cite{Parigi2015, Forbes2017}.
However, their characterization still relies on bulky polarization optics \cite{Arnold2019, Arnold2022}.
Fast, accurate and compact polarization measurements would be crucial for exploiting vector beams in widespread applications.

In this article, we demonstrate compact, single-shot measurements of multiple polarizations by photonic machine learning.
The idea is to map the beam polarizations into a complex spatial distribution of intensity corresponding to a point in a high-dimensional feature space. Then, supervised learning extracts polarization information from the intensity data. 
The critical point is that we avoid projecting on a polarization basis and perform no direct operation on the polarization state. 
The method thus removes the need for polarization optics and engineered devices in polarization tomography. 
We report accurate measurements of various beams encoding multiple SOP, including the case in which the number of polarization modes 
 is unknown and inferred by the measurements. 
An unexpected outcome is the experimental evidence of the double descent, a phenomenon that is attracting attention in machine learning \cite{Belkin2019}. The double-descent increases the classification accuracy when the dimension of the feature space is large. 
This effect improves the observation fidelity of the multiple SOP and enables high-precision measurements.

\begin{figure*}[p] 
\centering
\vspace*{-0.5cm}
\hspace*{-0.2cm} 
\includegraphics[width=2.05\columnwidth]{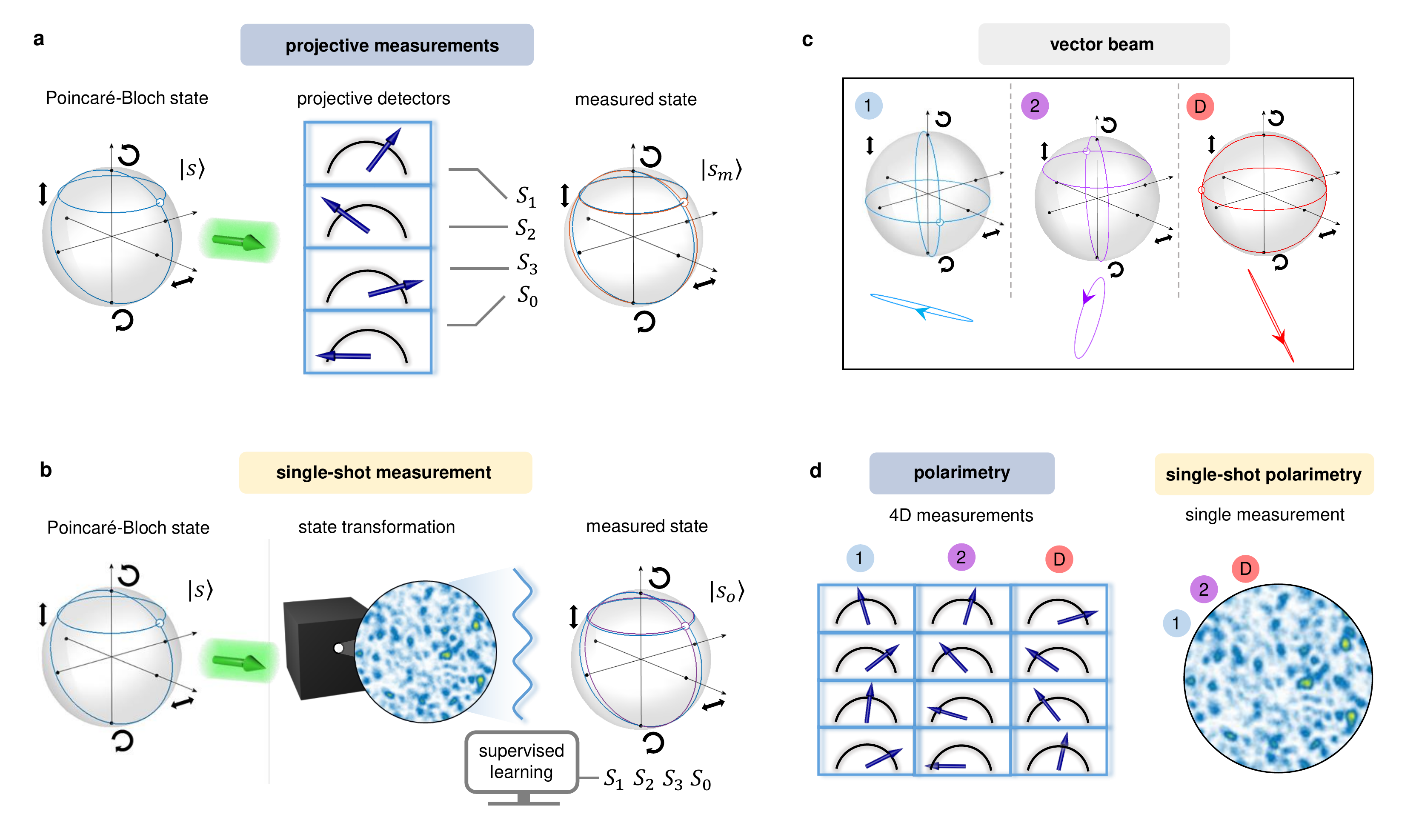} 
\vspace*{-0.4cm}
\caption{{\bf Observing multiple polarizations in a single shot.} (a) In projective measurements, the polarization state (a point on the Poincaré-Bloch sphere) is processed by distinct detectors, each analyzing one component. (b) In our single-shot measurement, the SOP is mapped onto a high-dimensional phase space (false-color map) where data are collected. The black box represents the transformation setup. The original polarization is identified in the feature space by supervised learning. (c)~Schematic of a vector beam composed of $D=3$ polarizations. 
(d) The projective analysis uses $4D$ measurements, while we use a single intensity distribution embedding all information. 
}
\vspace{-0.2cm}
\label{Figure1}
\end{figure*}
\begin{figure*}[p] 
\centering
\vspace*{-0.4cm}
\hspace*{-0.2cm} 
\includegraphics[width=2.05\columnwidth]{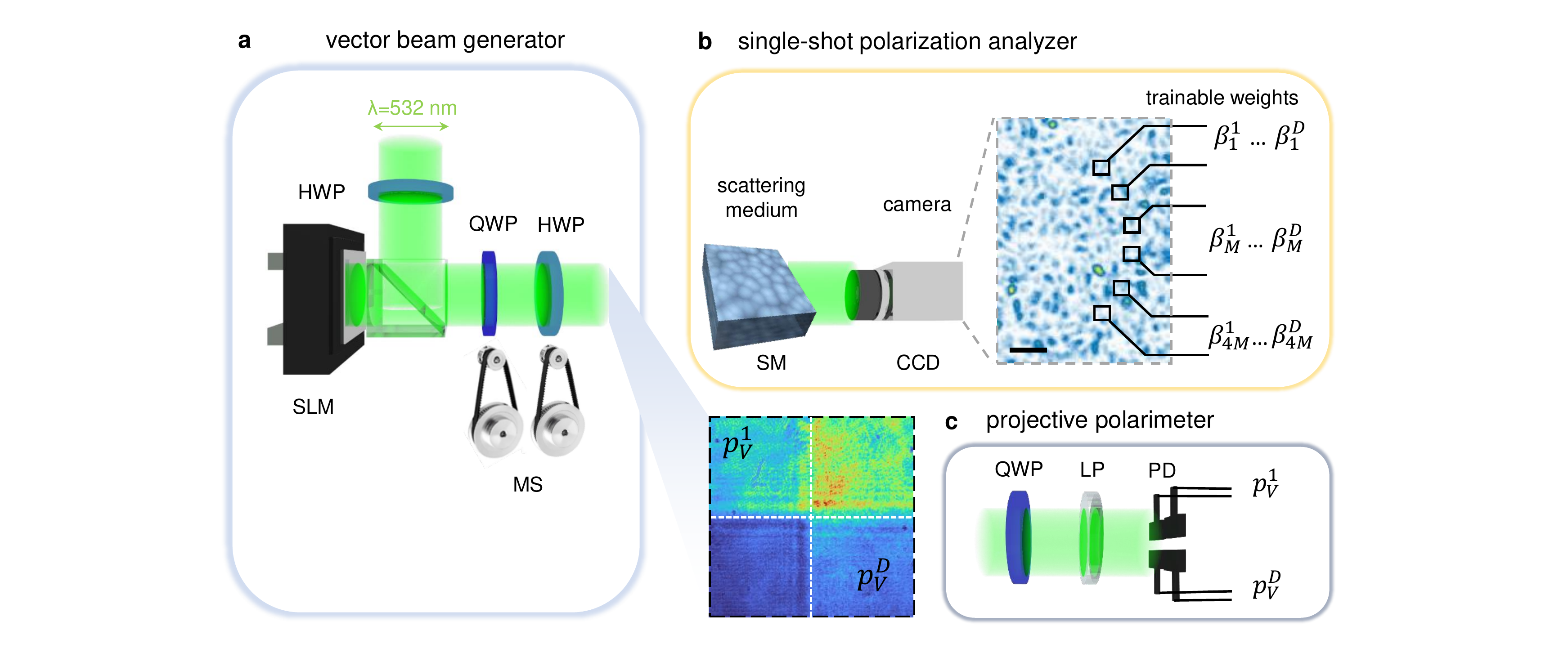} 
\vspace*{-0.3cm}
\caption{{\bf Experimental setup.} (a) Vector beams generator based on a phase-only spatial light modulator (SLM). 
Inset shows the vertical intensity component of a partitioned beam composed of four SOP. 
(b) Apparatus for single-shot polarimetry composed of a scattering medium (SM) and a camera (CCD).
The collected spatial intensity distribution (false-color inset) is processed on $4M$ linear output channels that enable training via the readout weights $\beta_i^j$. (c) Projective polarimeter with $D$ photodetectors (PD). HWP, half-wave plate. QWP, quarter-wave plate. LP, linear polarizer. MS, motor stages. 
}
\label{Figure2}
\end{figure*}
\begin{figure*}[t!]
\centering
\vspace*{-0.1cm}
\hspace*{-0.2cm}
\includegraphics[width=1.98\columnwidth]{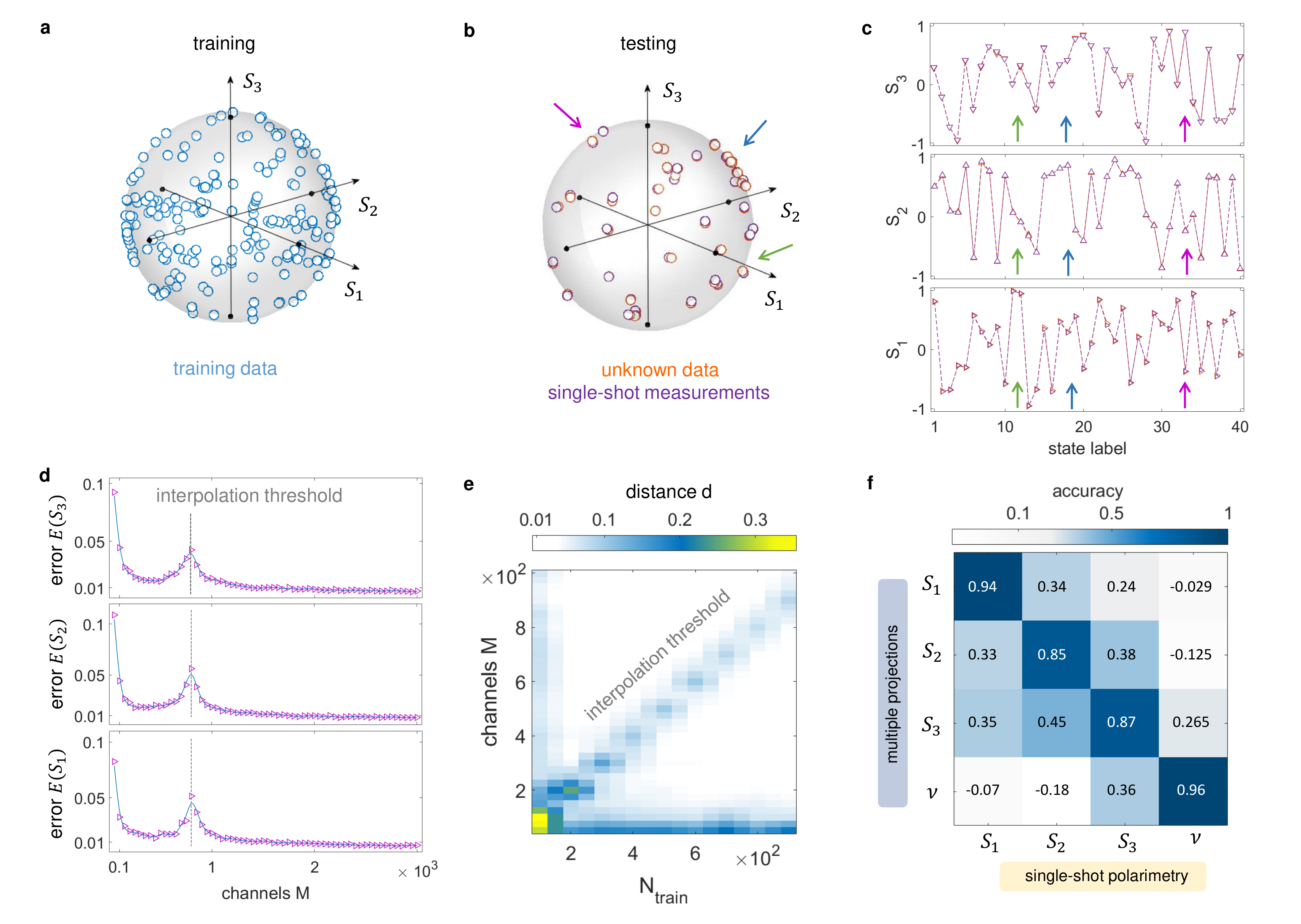} 
\vspace*{-0.2cm}
\caption{{\bf Polarization measurements by supervised learning.} 
(a) Training dataset made by $N_{\mathrm{train}}$ randomly-selected polarizations.
(b) Measured SOP and (c) corresponding Stokes parameters (purple), along with their unknown values used for testing (orange).
For clarity, only a subset of samples is shown. (d) Testing error $E(S_i)$ on each Stokes parameter versus the number of readout channels.
The error peak (interpolation threshold) is a direct evidence of the double-descent effect. 
(e) Distance between the measured and generated SOP varying the number of training samples and channels. 
(f) Accuracy matrix (see Methods) comparing single-shot ($M=200$) and conventional polarimetry (multiple projections).
}
\vspace{-0.1cm}
\label{Figure3}
\end{figure*}

\subsection{Single-shot polarimetry via scattering and learning}

Figure~1(a-b) reports our methodology. We show a representation of the SOP given by the four-component vector $\vert s \rangle = \vert S_1, S_2, S_3, S_0 \rangle$, with Stokes's parameters $S_i$. The system phase space is the unit sphere, known as the Poincaré-Bloch (PB) sphere. A projective measurement returns the components of the incident state $\vert s \rangle$ in a set of analysis vectors  [Fig.~1(a)]. Given four independent detected optical powers
 $\vert p \rangle = \vert p_1, p_2, p_3, p_4 \rangle$, the measured state is obtained as $\vert s_m \rangle=\hat{P}^{-1} \vert p \rangle$, where $\hat{P}$ is the instrument matrix \cite{Tyo2002, Kurtsiefer2006}. 

In our single-shot measurement [Fig. ~1(b)], the light beam interacts with a physical object that transforms
the input polarization into a set of features. The details of the object can be unknown to the observer.
The feature set at the readout is composed of many different observables. 
Specifically, we consider the field intensity at different $n$ spatial positions.
An intensity measurement collects the values  $\vert x \rangle = \vert x_1, x_2, ..., x_n \rangle$.
The polarization state is determined by a linear operation as
\begin{equation}
\vert s_o \rangle = \hat{\beta}  \vert x \rangle , 
\label{eq:1}
\end{equation}
where the $\hat{\beta}$ operator is a $4\times n$ matrix, which we refer to as the calibration matrix of the instrument.
Crucially, $\hat{\beta}$ is not determined {\it a-priori}, but retrieved by experimental data via machine learning.

The critical point of our single-shot method is the redundant mapping of a polarization state (determined by four observables) to a state belonging to a much larger space, defined by $n$ observables. If $n$ is very large, the redundancy makes the method very advantageous to measure light beams
that encodes many polarizations in distinct spatial optical modes [Fig.1(c)]. 
The corresponding multiple SOP can be realized by $D$ individual polarization states as 
$ \vert \bold{s} \rangle = \vert s^1 \rangle \oplus \vert s^2 \rangle \oplus ... \oplus \vert s^D \rangle$, 
where $\vert s^j \rangle $ denotes the Stokes vector of the $j$-th mode.
The observation of multiple SOP with conventional polarization tomography requires at least $4D$ projections 
or $4D$ generalized measurements \cite{Arnold2022}.
On the contrary, we observe the state $ \vert \bold{s}_o \rangle$, composed by $D$ individual SOP,
by a single-shot measurement by using the higher-dimensional data vector $\vert x \rangle$.
The needed calibration matrix $\hat{\beta}$, with size $4D \times n$, is determined by an initial training phase.
The scheme is scalable with the polarization state dimension $D$, at variance with projective measurements, where an additional detector is required for any dimension. In our case, we can observe input states $\vert \bold{s} \rangle$ of variable dimension
by adopting the same feature space. 
This property means we can obtain additional beam parameters otherwise complex to access through the same detected signal.

\begin{figure*}[t!]
\centering
\vspace*{-0.1cm}
\hspace*{-0.2cm}
\includegraphics[width=1.85\columnwidth]{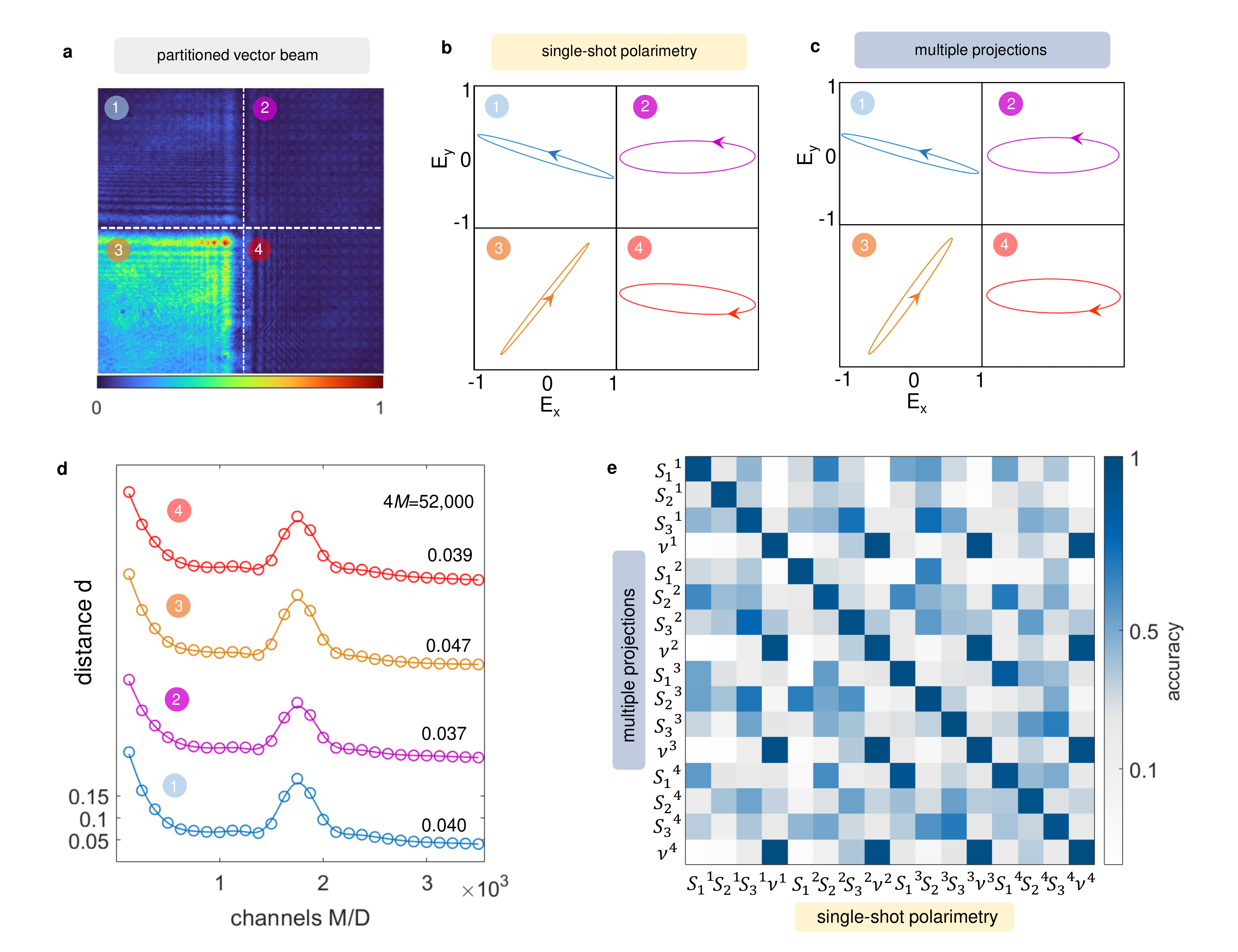} 
\vspace*{-0.0cm}
\caption{ {\bf Single-shot polarimetry of four-parted vector beams.}
(a)~A generated beam with $D=4$. We show the spatially-resolved vertical intensity component $p_V$. 
(b)~Polarization ellipses observed in a single shot ($M/D=1000$) and (c) via multiple projective measurements. 
(d)~Distance of the $j$-th mode as a function of the channel number $M$. Data are vertically shifted for clarity; 
inset values are the minimum $d$ for each curve. (e) Accuracy matrix of the single-shot polarimetry for $D=4$.
} 
\vspace{-0.0cm}
\label{Figure4}
\end{figure*}

Single-shot polarimetry is experimentally implemented by using the scheme illustrated in Fig.~2(a-b).
The optical setup is composed of two parts: a generator that produces vector beams
and a single-shot polarization analyzer that realizes the optical transformation and collects the resulting intensity (details in Methods). 
To generate multiple SOP we exploit a phase-only spatial light modulator (SLM). This allows shaping a large number of SOP on the wavefront of a single $532$ nm laser beam, with distinct states $\vert s^j \rangle $ that correspond to spatially-separated modes with phase $\phi_j$ (see Supplementary Note 1). The single-shot analyzer is based on multiple light scattering~\cite{Popoff2010}. We make use of a glass diffuser to map the SOP
into intensity data. However, a similar mapping can be performed by any optical system sensitive to the input polarization.
The scattering medium spatially mixes the incoming optical field and transmits a disordered intensity distribution.
The resulting speckle pattern is imaged on a camera sensor with no polarization filters. 
The scatterer couples the polarization and spatial amplitude degrees of freedom.
However, this coupling is very small, i.e., for a uniform beam, the light arriving on the camera has a degree of polarization close to one.
This property corresponds to specific correlations in the medium vectorial transmission matrix (Supplementary Note 3). 
Thus, we get a transmitted intensity image in which the spatial details depend on the vector beam polarizations (Supplementary Note~2).
By sampling the image, we obtain the output data $\vert x \rangle$. This vector encodes the input polarizations into a higher-dimensional feature space.

To extract polarization information from the intensity data, we use supervised learning.
Among the various neural network architectures that have been proved convenient for optical systems~\cite{Psaltis2018}, 
we adopt the extreme learning machine (ELM)~\cite{Huang2012}. ELM allows fast and easy training with several thousands of network nodes,
thus being especially suited to large-scale photonic implementations \cite{Wetzstein2020, Saade2016, Marcucci2020, Pierangeli2021, Rafayelyan2020, Psaltis2021, Ciuti2022, Valensise2022}.
To apply the ELM algorithm, we construct a readout layer by a random selection of $4M$ output camera channels [Fig. 2(b)]. 
Each channel has a linear weight $\beta_i^j$, with $i=1,...,4M$ and $j=1,...,D$, which form the calibration matrix in Eq. (1).
Training consists in adjusting these parameters according to a labeled dataset via ridge regression (see Methods).

\begin{figure*}[t!]
\centering
\vspace*{+0.0cm}
\hspace*{-0.6cm}
\includegraphics[width=2.18\columnwidth]{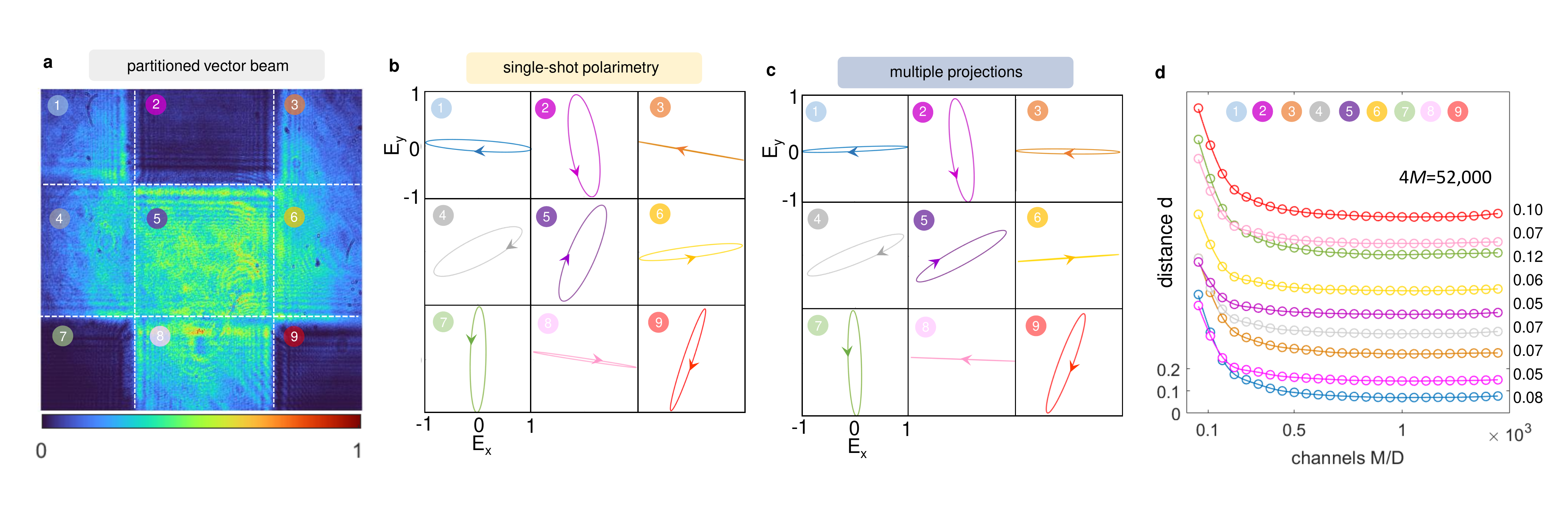} 
\vspace*{-0.4cm}
\caption{{\bf Single-shot polarimetry of nine-parted vector beams.} 
(a)~Intensity $p_V$ observed when vertical projecting a generated beam with $D=9$.
The total state $ \vert \bold{s} \rangle $ belongs to a $27$-dimensional phase space.
(b)~Single-shot measurement of the nine polarizations. 
(c)~Polarization tomography from multiple measurements for comparison.
(d)~Distance between the single-shot observations and generated SOP. 
Curves are shifted for visualization; the inset shows their value for $52,000$ channels.
}
\vspace{+0.2cm}
\label{Figure5}
\end{figure*}
\begin{figure}[t!]
\centering
\vspace*{-0.4cm}
\hspace*{-0.4cm}
\includegraphics[width=1.05\columnwidth]{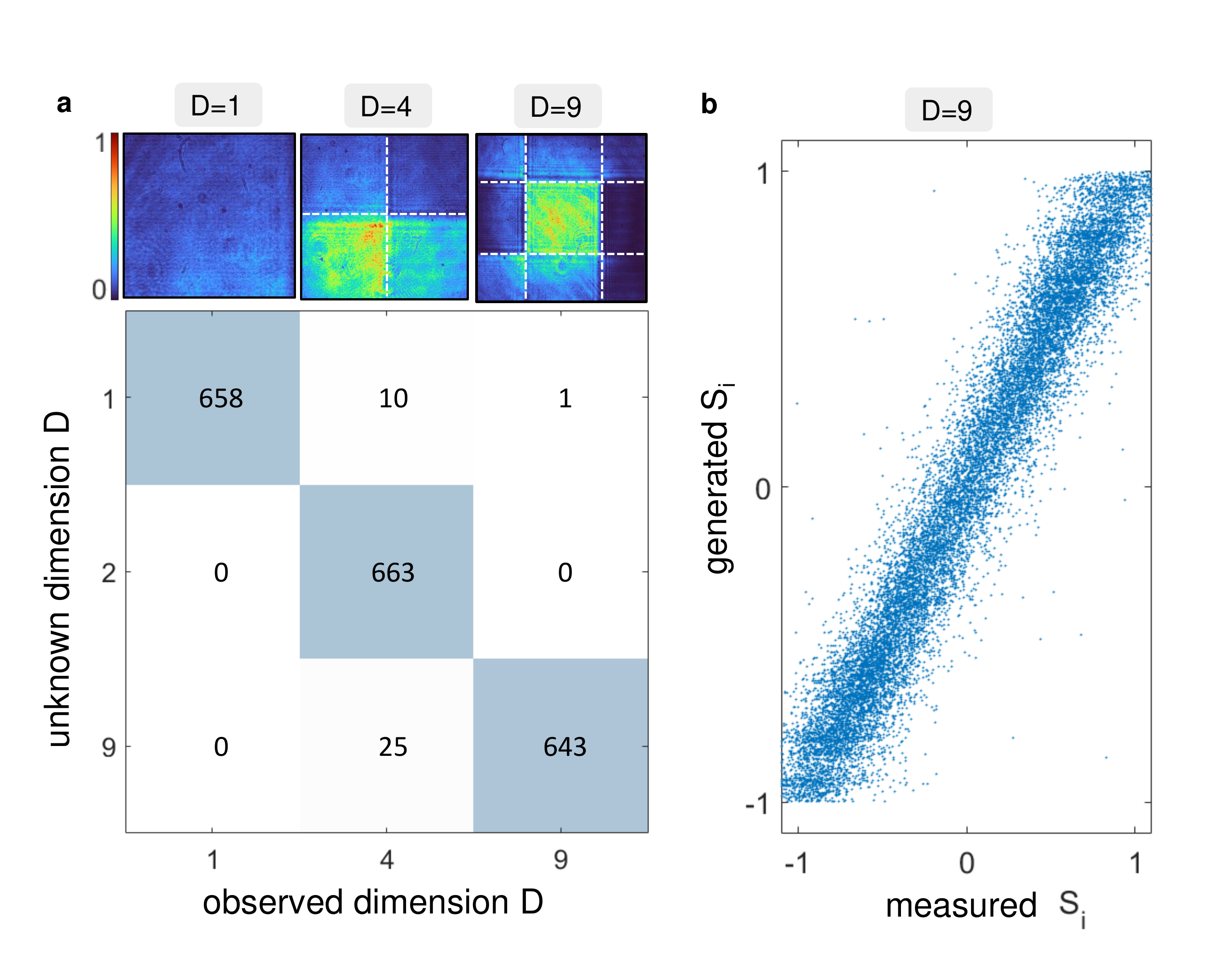} 
\vspace*{-0.4cm}
\caption{ {\bf Measurement of partitioned beams of unknown polarization number.} 
(a) Confusion matrix over $N_{\textrm{test}}=2000$ beams with three possible values for $D$.
Insets are representative projective images in the three cases. $D$ is determined with $98$\% accuracy. 
(b) Single-shot measured vs generated Stokes parameters for beams of nine part, whose dimension has been automatically identified.
} 
\vspace{-0.1cm}
\label{Figure6}
\end{figure}

Figure~3 shows single-shot polarimetry of a single SOP. We generate $N_{\mathrm{train}}$ samples randomly distributed on the PB sphere [Fig.~3(a)]
and validate the calibrated single-shot analyzer on $N_{\mathrm{test}}$ unknown SOP (Methods). 
As reported in Fig.~3(b), the observed polarizations correspond to the input within a distance $d=0.014\pm 0.02$ (Methods).
Each measured Stokes parameter $S_i$ is in remarkable agreement with its target value [Fig.~3(c)].
The accuracy of the single-shot polarimeter is investigated by varying the number of network channels $M$
and the size $N_{\mathrm{train}}$ of the calibration dataset. Results in Fig.~3(d) show the behavior of the error $E(S_i)$. 
We find that single-shot measurements become extremely accurate when $M$ increases [$E(S_i) = 0.0075$].

The observed error peak in Fig.~3(d) discloses the so-called double descent~\cite{Belkin2019}.
Above the interpolation threshold, the accuracy increases with the number of channels with no overfitting. 
We find that the interpolation threshold shifts with the training dataset size [Fig.~3(d)], 
with a maximum error for $M \simeq N_{\mathrm{train}}$. 
This point marks a critical condition in ELM with physical systems~\cite{Marcucci2020}.
Experimental observation of the double descent is made possible by the huge number of output nodes of our optical setup.
On the other hand, we underline that a few hundred channels are sufficient for effective polarization measurements.
Figure~3(f) shows the accuracy matrix $\hat{a}$, which also includes the degree of polarization $\nu$ (see Methods).
The single-shot results match well data from a conventional rotating-waveplate polarimeter [Fig.~2(c)]. 
With reference to single-qubit tomography~\cite{Altepeter2005}, 
we obtain a fidelity $F(\rho_s, \rho_m)= 0.99 \pm 0.01$, where $\rho_s$ and $\rho_m$ are the density matrices obtained via single-shot and multiple projections.

\subsection{Single-shot polarimetry of vector beams}

In Figure 4 we report single-shot polarization measurements of partitioned vector beams with four SOP.
Calibration operates alike the single polarization case, but with $N_{\mathrm{train}}$ states with $D=4$.
Figure~4(a) shows the intensity of an unknown SOP generated by four random phases $\phi_i$ on four spatial modes. 
The effect of diffraction between adjacent SOP is visible.
By collecting the transformed intensity image, we reveal the full polarization structure in a single shot, 
as shown by the polarization ellipses in Fig.~4(b). 
Conventional projective polarization tomography strikingly agrees with the single-shot observation [Fig. 4(b-c)].  
The overall uncertainty of the detection for $D=4$ is evaluated in Fig.~4(d) by varying $M$ up to $4M=52,000$ total channels. 
The error keeps decreasing with the size of the signals, with localized peaks that indicate the interpolation threshold. 
The fidelity is confirmed by the overlap matrix $\hat{a}$ in Fig.~4(e), with $\mathrm{Tr} \left(  \hat{a}\right)/4D =0.95$ accuracy.

We succeed in characterizing with high accuracy beams encoding up to nine SOP. 
This implies a single-shot reconstruction of a classical state $\vert \bold{s} \rangle$ that lye in a $27$-dimensional phase space.
To implement larger systems we expand the spatial extent of the input and output optical planes. Figure~5a shows a multiple  SOP with $D=9$.
The result of the single-shot detection agrees with the state from multiple projections [Fig.~5(b-c)].
The comparison, with the $\hat{a}$ matrix having $1296$ entries, gives a $0.91$ overlap between the polarization states measured with the two methods. 
The single-shot observation is hence performed with high precision, as also the measurement error in Fig.~5(d) indicates. 
Interestingly, we observe that the distance $d$ decreases rapidly with $M$ and gets stuck into a broad plateau.
This behavior underlines the enormous complexity of the states that we are characterizing. 
Even if in Fig.~5(d) we do not observe any interpolation threshold as even more output channels would be necessary, 
the method still enables accurate measurements.

\vspace*{-0.1cm}
\subsection{Measuring the number of polarizations}

In the most general case of an unknown multiple-polarization state, its dimension $D$ is also an unknown variable.
We demonstrate that we can identify also the dimensionality of the multiple SOP. 
To train the setup, the calibration dataset includes multiple SOP with a 
variable number of partitions, and $D$ as an additional target parameter (Methods).
Figure~6(a) reports the confusion matrix obtained for vector beams having three possible partition configuration 
($D=1$,$D=4$, and $D=9$), with $M=10000$. The dimension is found with an accuracy that exceeds 98\%.
In addition, the set of Stokes parameters is measured with precision comparable to cases of known dimension [$E(S_i) = 0.057$]. 
For $D=9$, a comparison between the entire $S_i$ distribution is shown in Fig.~6(b).
Therefore, we not only perform single-shot polarimetry of beams encoding nine polarizations,
but we carry out the measurement not knowing that the SOP has nine dimensions.
It is important to note that the dimension $D$ is not directly measurable by using multiple projective measurements.

\vspace*{-0.2cm}
\subsection{Conclusions}

We have experimentally demonstrated the measurement of multiple polarizations in a single shot without polarization optics.
The result has been obtained with an original method that combines physical transformation between the polarization and spatial degrees of freedom
with machine learning to get unprecedented information in a single detection.
Our measurement scheme has no bulky optical components and can operate at any wavelength.
The single-shot polarimeter is hence compact, without moving components, and does not require nanofabrication. 
Moreover, it provides direct access to properties of the vector field otherwise difficult to quantify,
as we demonstrate by measuring the unknown number of polarizations within the vector beam.

These findings empower compact single-shot polarimetry based on machine learning in a wide variety of contexts,
from optical networking to biomedical devices. We foresee the extension of our single-shot approach to the entire electromagnetic spectrum \cite{Gongora2022}, 
to subwavelength and topological optical fields, and, more generally, to other optical degrees of freedom~\cite{Aiello2015_2}, 
with applications where conventional instruments are useless, as in edge devices and photonic chips~\cite{Fratalocchi2019}.
Moreover, partitioned vector beams of quantum light can encode many qubits.
Therefore, our results may also open exciting perspectives in the quantum domain, 
with the possibility to benefit from single-shot polarimetry by machine learning.

\vspace*{+0.1cm}
\noindent {\bf Acknowledgments.} We acknowledge funding from the Italian Ministry of University and Research (PRIN PELM 20177PSCKT),
Sapienza Research, QuantERA ERA-NET Co-fund (Grant No. 731473, project QUOMPLEX), and H2020 PhoQus project (Grant No. 820392). 
We thank L. Dieli for useful discussions, I. MD Deen and F. Farrelly for technical support in the laboratory. 

\noindent {\bf Author contribution.} D.P. and C.C. developed the idea. D.P. carried out experiments and data analysis.

\clearpage
\newpage

\section{Methods}
\vspace*{+0.0cm}
{\small 

{\bf Experimental setup.} The experimental setup follows the sketch in Fig. 2.
A continuous-wave laser beam with wavelength $\lambda= 532$nm (LaserQuantum Ventus 532, 250 mW) 
is expanded and linearly polarized along the horizontal (H) direction ($x$-axis) with a linear polarizer (LP).
The vector beam generator is composed of a reflective phase-only SLM
(Hamamatsu X13138, $1280\times1024$ pixels, $12.5 \mu$m pixel pitch, $60$Hz frame rate)
sandwiched between an input half-wave plate (HWP) and an output polarization system made by a quarter-wave plate (QWP) and HWP.
The output waveplates are equipped with motorized precision rotation stages (MS) ($25$\textdegree /sec maximum rotation velocity)
that are programmed on-line. Their fast axes form respectively an angle $\alpha$ and $\beta$ with the $x$-axis.
The parameters $\alpha$ and $\beta$ are varied with $2$\textdegree resolution during both training and testing.
By grouping $L \times L$ SLM pixels, the modulator active area is divided into $D$ squared input modes,
with each mode having a phase $\phi_j$ in the $[0, 2\pi]$ interval. The available phase levels are 
$210$, distributed according to a linear phase response curve \cite{Ruan2020}.
Polarization-modulated light is focused by a plano-convex lens (f$=150$mm) on a scattering medium 
(Thorlabs N-BK7 Ground Glass Diffusers, 1500 grit) positioned using a four-axis translational stage. 
The scattered field is collected by an imaging objective (NA$=0.25$) and the transmitted speckle pattern
is detected by a cooled camera (Basler a2A1920-160umPRO, $1920\times1200$ pixels, $160$fps) with $12$-bit (4096 gray-levels)
intensity sensitivity.
Within the camera region of interest, $4M$ output channels are randomly pre-selected. The signal is obtained by binning 
over a few camera pixels ($4\times4$) to reduce detection noise. Output channels have a size comparable with the spatial extent of a speckle grain.
For reference measurements, a portion of the polarization-modulated beam is split and analyzed with conventional polarimetry.
A commercial rotating-waveplate polarimeter (PAX1000VIS, $0.25$\textdegree nominal accuracy) is used for single SOP. 
For projective polarization imaging of vector beams, we use a custom rotating-waveplate analyzer composed of a QWP, a LP, a power meter (PM100D), 
and a CMOS camera (DCC1545M). Intensity projections are measured along the horizontal (H), vertical (V), diagonal (D) and right circular (R) component,
i.e., $p_H$, $p_V$, $p_D$, $p_R$ are the detected power. An example of projective analysis for a partitioned vector beam 
is reported in Supplementary Fig. 2.

\vspace*{0.2cm}
{\bf Training method.} 
Calibration of the setup for single-shot polarization measurements is performed by generating $N_{\mathrm{train}}$ multiple SOP 
and by loading one-by-one the corresponding phase mask on the SLM.
For the $l$-th training sample, the output waveplates of the vector beam generator are rotated to the 
couple of parameters $\alpha_l$ and $\beta_l$ (Supplementary Note 1). 
The dataset $\lbrace\phi^j_l, \alpha_l, \beta_l\rbrace$ is randomly generated to cover the entire phase space.
When the polarization dimension $D$ need also to be determined, the training set is composed of states with a variable number of polarizations.
Intensity values from the $4M$ output channels are stored (Supplementary Fig. 6).
We increase linearly the number of selected channels when increasing the dimension, $M\propto D$.
In the case of $D=9$, for the feasibility of training, we consider a maximum value of $M=13000$.
For the measurement of SOP with unknown polarization number, $M$ additional channels are used to reconstruct $D$.
The calibration weights $\beta_i^j$ are determined by applying the training algorithm to the entire set of acquired data.
The obtained calibration matrix $\hat{\beta}$ is used to measure $N_{\mathrm{test}}$ random SOP.
We use a ratio $N_{\mathrm{test}} / N_{\mathrm{train}} = 0.1$.

\vspace*{0.2cm}
{\bf Extracting polarization information from intensity data.}
In our setup, the scattering medium creates a mapping between the incoming polarization set and a higher-dimensional feature space.
This general mechanism grounds physical implementations of ELM and kernel machines for neuromorphic computing \cite{Saade2016, Pierangeli2021, Psaltis2021}. In our case, the scheme is trained to perform physical measurements.
To determine the calibration matrix $\hat{\beta}$ by using $4M$ output channels, we consider a training set of randomly-selected Stokes vectors 
$\lbrace\vert \bold{s} \rangle \rbrace \equiv \bold{S}$ as target vectors ($N_{\mathrm{train}} \times 4D$ sized).
The corresponding acquired intensity matrix
is $\lbrace \vert \bold{x} \rangle \rbrace \equiv \bold{X}$, with size $N_{\mathrm{train}} \times 4M$ ($M \gg D$).
Training corresponds to solving numerically the ridge regression problem: 
\begin{equation}
\argminG_{\hat{\beta}} ( \Vert \mathbf{X}\hat{\beta} - \mathbf{S} \Vert^2 + c^{-1} \Vert \hat{\beta} \Vert^2),
\end{equation}
where the parameter $c$ controls the trade-off between the training error and the regularization.
A solution is given by~\cite{Huang2012}
\begin{equation}
\hat{\beta}= \left( \mathbf{X}^{T}\mathbf{X} +  c \mathbf{I} \right)^{-1} \mathbf{X}^{T} \bold{S} ,
\end{equation}
where $\mathbf{I}$ is the identity matrix. Inversion involves the $4M \times 4M$ matrix $\mathbf{X}^{T}\mathbf{X}$, 
which makes the method scalable as $M$ is selected by the experimenter.
Therefore, given an unknown multiple SOP with the single-shot intensities $\vert x \rangle$, 
the observed state can be expressed as 
\begin{equation}
\vert \bold{s}_o \rangle =  \left( \mathbf{X}^{T}\mathbf{X} +  c \mathbf{I} \right)^{-1} \mathbf{X}^{T} \mathbf{S} \vert x \rangle .
\end{equation}
An explicit expression for $x_i$ is reported in Supplementary Note 2.
In the case of a single SOP, the measured Stokes's parameters from single-shot intensity data are
\begin{equation}
S_i = \sum\nolimits_{k=1}^M \beta_k x_k.
\end{equation}
Each $S_i$ is hence decoded through $M$ weights. An example of both the acquired intensity and calibration matrix is
reported in Supplementary Note 4.

\vspace*{0.2cm}
{\bf Analysis of the measurements.}
To quantify the accuracy of the single-shot polarimeter, as a testing error on the $i$-th Stoke's parameter
we use the mean-absolute-error (MAE)  $E(S_i)=\langle \vert S_i^s - S_i^g \vert  \rangle$, 
in which the apex stands for single-shot (s) and generated (g), and the average is over $N_{\mathrm{test}}$ samples.
Instead of $S_0$, we report the degree of polarization $\nu=\sqrt{S_1^2 + S_2^2 + S_3^2}/S_0$,
which quantifies the amount of unpolarized light.
The distance between two single SOP on the PB sphere is computed as $d= \sqrt{\sum_i E(S_i)^2}$.
The accuracy values with respect to projective measurements are $a_{ij}=1 - O_{ij}$, with the overlap
$O_{ij}=\langle \vert S_i^s - S_i^m \vert \rangle$, with apex $m$ stands for multiple measurements.
It is worth noting that this comparison includes also the inaccuracy of the SOP generator and the uncertainty
of the conventional polarization analyzer (Supplementary Fig. 2).
For the single SOP, the fidelity is computed in analogy with quantum state tomography as
$F(\rho_s, \rho_m)= \mathrm{Tr} \left( \sqrt{ \sqrt{\rho_m} \rho_s  \sqrt{\rho_m}  } \right)^2 $,
where the density matrix is $\rho = 1/2 (\sum_i S_i \hat{\sigma_i})$ with Pauli matrices~$\hat{\sigma_i}$.

}


\end{document}